# Superfluorescent upconversion nanoparticles as an emerging second generation quantum technology material

**Lewis E. MacKenzie**[†*] and **Peter Kirton**[‡‡]

[†]Department of Pure and Applied Chemistry, University of Strathclyde, Technology Innovation Centre, 99 George Street, Glasgow, Scotland, United Kingdom (UK), G1 1RD. * Corresponding Author.
Email: l.mackenzie@strath.ac.uk

[‡‡]Department of Physics and Scottish Universities Physics Alliance (SUPA), University of Strathclyde, Glasgow, G4 0NG, United Kingdom.
Email: peter.kirton@strath.ac.uk

**Abstract**

Superfluorescence (SF) in lanthanide doped upconversion nanoparticles (UCNPs) is a room-temperature quantum phenomenon, first discovered in 2022. In a SF process, the many emissive lanthanide ions within a single UCNP are coherently coupled by an ultra-short (ns or fs) high-power excitation laser pulse. This leads to a superposition of excited emissive states which decrease the emissive lifetime of the UCNP by a factor proportional to the square of the number of lanthanide ions which are coherently coupled. This results in a dramatic decrease in UCNP emission lifetime from the μs regime to the ns regime. Thus SF offers a tantalizing prospect to achieving superior upconversion photon flux in upconversion materials, with potential applications such as imaging and sensing. This perspective article contextualizes how SF-UCNPs can be regarded as a second generation quantum technology, and notes several challenges, opportunities, and open questions for the development of SF-UCNPs.

## 1. Introduction

We are living in the age of quantum technologies. First-generation quantum technologies can be generally defined as arising from 1940-1990 and were based upon the principals of particle spin (e.g. NMR, spintronics) and quantum tunnelling (e.g. scanning tunnelling microscopy, tunnelling diodes, Josephon junctions (superconductors), and superconducting quantum interference devices (SQUIDs)).

A second generation quantum technology can be succinctly defined as any technology in which one or more quantum processes — e.g. quantum entanglement, coherence, confinement, spin, etc. — is exploited for the purposes of a second generation quantum technology application, e.g. quantum sensing, quantum imaging, quantum timing, quantum navigation, quantum cryptography, or quantum computing (see Figure 1).[1–12] Second generation quantum technologies generally tend to have been developed after 1990.[5] Accompanying these second generation quantum technologies are supporting technologies, such as cooling systems, electronics, production technologies, and optical control systems.[6,13]

It is arguable that some optical materials may be regarded as components, or even the core part of a quantum technology. For example, quantum dots (QDs) operating on quantum confinement principals[14] are arguably a first generation quantum technology (i.e. pre-1990s),[5] which amongst their myriad applications in displays, imaging, and sensing, have subsequently been applied to various second generation quantum technology application areas, e.g. quantum simulation, and quantum computing.[6,15–19]

Photonic upconversion materials were first developed in bulk form in the 1960s,[20] with a subsequent eruption of scientific interest in upconversion nanoparticles (UCNPs) since the early 2000s.[21,22] In brief, upconversion materials consist of a crystalline host structure, such as the commonly utilized sodium yttrium fluoride ($NaYF_4$).[23] In this host lattice are various lanthanide ions which can absorb multiple low-energy photons, such as neodymium, ytterbium, and erbium ions ($Nd^{3+}$, $Yb^{3+}$, and $Er^{3+}$) which absorb at ~800 nm, ~980 nm, and ~1500 nm respectively. These sensitizer ions then non-radiatively transfer energy to various emissive ions, such as erbium and thulium ($Er^{3+}$, $Tm^{3+}$) resulting in emission of a higher-energy photon, typically in the visible wave range.[23–26] Hence the name "upconversion", deriving from the upconversion of photon energy.

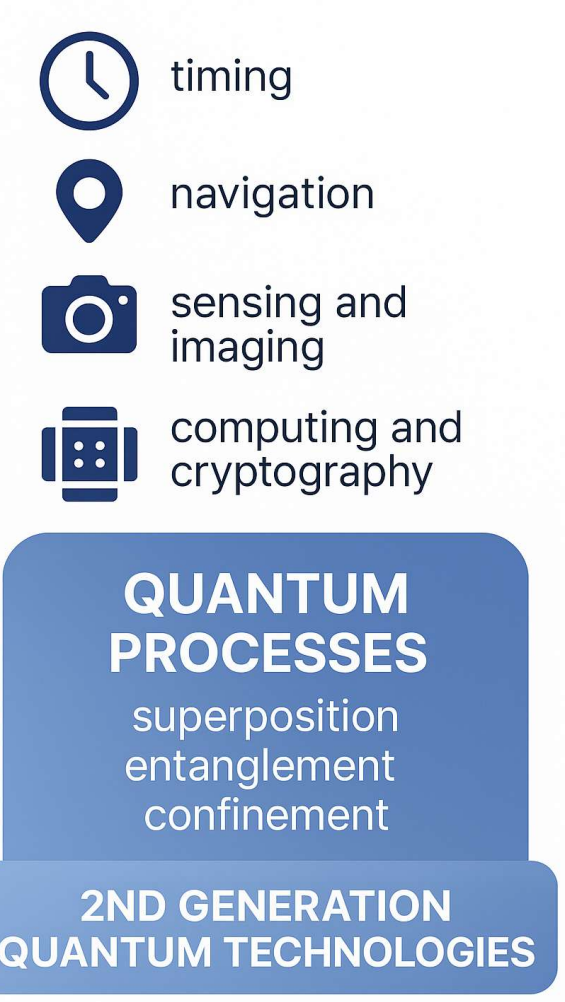

**Figure 1.** Overview of key quantum processes and applications for 2[nd] generation quantum technologies.

UCNPs have many attractive practical photonic properties, such as near-infrared and infrared excitation, resistance to photobleaching, physically robust nature, and tunable "line-like" emission. Additionally, their long-lived emission lifetime (typically $\tau \sim 100$ - 1000 μs arising from $4f \rightarrow 4f$ parity-forbidden lanthanide transitions),[27,28] can be utilized for time-gated imaging and sensing.[29,30] Consequently, UCNPs have attracted widespread interest in applications such as temperature sensing,[31] pressure sensing,[31,32] plasmonics,[33] photocatalysis,[34] solar cells,[35] security inks,[29,36] data storage,[37,38] display screens,[39] biosensing,[40] photodynamic therapy, multi-modal deep-tissue biological imaging,[30,41] and optogenetics.[42,43]

Whilst exhibiting many desirable optical properties, UCNPs have the downside of being relatively inefficient optical emitters, with the highest UCNP quantum yield on record being ~9% for core-shell $NaYF_4$:Yb,Er@$NaYF_4$ beta-phase UCNPs in dry form.[23] As a colloidal suspension in organic solvents or water, the quantum yield of UCNPs is more typically fractions of a percent.[44,45] In particular, UCNPs are prone to quenching by the OH- vibrations in water molecules.[28] In contrast quantum dots (QDs) offer quantum yields in ranges of 25-75% and rhodamine 6G in ethanol has a quantum yield of 95%.[46,47] This relatively inefficient emission has appeared to be a fundamental limitation of upconversion materials for decades.

Despite the lanthanide-based photophysics of UCNPs being underpinned and described by quantum processes,[48] UCNPs themselves have typically not meet the criteria to be considered an optical material for second generation quantum technologies due to lack of appropriate quantum processes and appropriate applications. This is in contrast to related materials, such as polynuclear lanthanide complexes and solid rare earth ion systems, which have been proposed as quantum bits (qubits) for quantum information processing and quantum computing.[49–51]

Recent demonstrations of a new quantum phenomenon in UCNPs — superfluorescence (SF), also sometimes referred to as Dicke superradiance, — has changed how we can consider UCNPs to be quantum optical nanomaterials. SF in UCNPs was first demonstrated by a group of researchers in the USA (Huang et al., 2022),[52] where the SF phenomenon resulted in $Nd^{3+}$ doped UCNPs emitting photons at a rate ~10,000 greater than via conventional upconversion emission processes. Notably SF occurs at room-temperature.[52] A second study has demonstrated that SF can achieve 70x enhancement in UCNP emission intensity.[53] Therefore SF-UCNPs bypass the fundamental limitations of UCNPs, and SF-UCNPs could be exploited as a second generation quantum technology in any application where an increased photon emission rate and corresponding increase in brightness is advantageous, with obvious utility in sensing and imaging technologies.

**2. Superfluorescence in optical materials**

SF is a quantum optical phenomena based upon quantum coherence (i.e. the phase stability of a superposition state).[54] In SF systems, individual emissive dipoles (normally of a random phase) are coherently coupled to generate a single (phase aligned) large collective macroscopic dipole, resulting in very fast emission (see Figure 2). This is analogous to a large crowd at a sporting match: normally what any individual is saying cannot be heard clearly – but in the aligned state, i.e. the crowd chanting together (analogous to SF) – the message is very clear! [55]

SF was first proposed by Dicke in 1954,[56,57] and was first experimentally demonstrated in 1973.[58] In typical SF systems, the SF lifetime is proportional to normal emission lifetime of the system divided by the number of phase aligned dipoles, whereas SF emission intensity scales proportionally to the square of the number of aligned dipoles (see Figure 2).[52] It has been proposed that SF systems may be of utility for quantum computing, cryptography,[55] and super-resolution microscopy.[53]

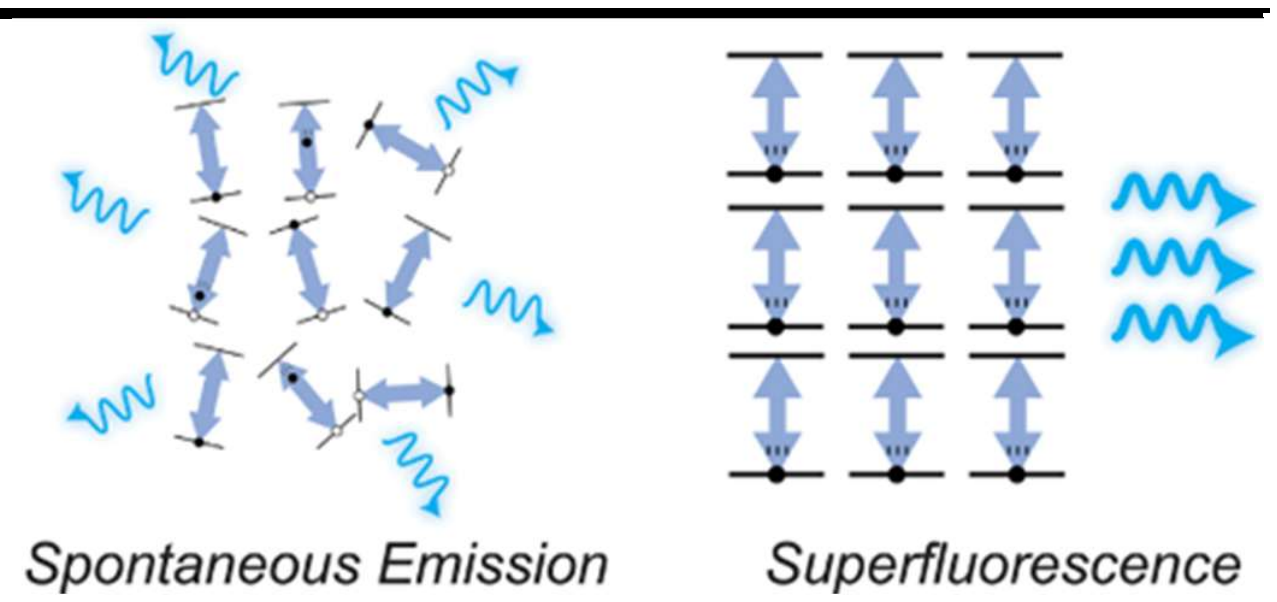

**Figure 2.** Simple diagram depicting SF. Left: under conventional emission conditions, the phase of each individual dipole emitter (represented by direction of arrow) is randomly aligned. Right: under SF conditions, the phase of many individual dipole emitters is aligned, via coherence coupling and so the system acts as a macroscopic giant dipole. *Figure reproduced in part from Russ and Eisler (2024)[55] under a Creative Commons 4.0 International Licence.*

The key concept underpinning the SF effect is that of constructive interference between emission from the dipoles.[56,59] As long as they are separated by less than the emission wavelength, then emission from each individual dipole is indistinguishable and the emission process descends the Dicke ladder of states. For an example of an ensemble of 3 dipoles, which can either be in the excited ($e$) or ground ($g$) state, this process looks like the following:

| | |
|---|---|
| $\lvert eee\rangle$ <br> $\Downarrow$ | All dipoles are in their excited state. ($N = 3$) |
| $eeg\rangle + \lvert ege\rangle + \lvert gee\rangle$ <br> $\Downarrow$ | Intermediate superposition of all possibilities with one dipole in its ground state. |
| $\lvert egg\rangle + \lvert geg\rangle + \lvert gge\rangle$ <br> $\Downarrow$ | Intermediate superposition of all possibilities with two dipoles in the ground state. |
| $\lvert ggg\rangle$ | All dipoles are in the ground state. |

Where the notation $|ijk\rangle$ denotes the wavefunction with the dipoles in the states $i$, $j$ and $k$. Since we cannot identify exactly which dipole has decayed, we end up with the superposition states indicated. This leads to the behavior that in the middle of this ladder of states, many decay pathways interfere constructively with each other, leading to an enhanced rate of emission.

To elaborate further, the Dicke states are eigenstates of the total angular momentum operator for the ensemble of dipoles, and so can be labelled just by the number of excited dipoles, $n$. These states are shown schematically in Figure 3. The equation describing the number of excited dipoles, $n$, is:

$$\frac{dn}{dt} = \gamma n(N - n) = \Gamma(n). \qquad \text{Equation 1}$$

Where: $\gamma$ is the excited state lifetime, $N$ is the number of indistinguishable dipoles (within a volume covered by the emission wavelength), and $\Gamma$ is the collective emission rate.[59] We see that in the middle of the emission process, i.e. when $n \sim \frac{N}{2}$ (i.e. that half the initial number of excited dipoles remain in an excited state), that the rate of emission, and hence the intensity of light emitted, is proportional the square of the initial number of coupled dipoles, $N^2$. Therefore to maximize the SF emission rate (i.e. minimize SF emission lifetime), it is desirable to maximize the initial number of coherently coupled dipoles.

In contrast, in the case of independent dipoles which are not coherently coupled (i.e. non-SF state), the emission rate is always proportional to $N$ (see Figure 3).

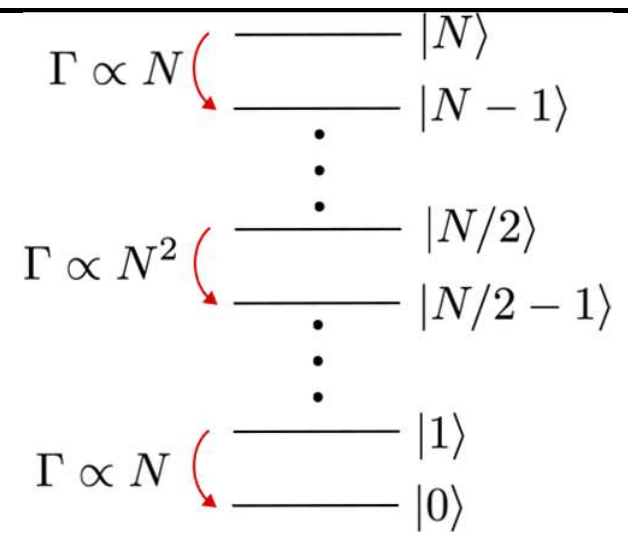

**Figure 3.** Energy level schematic showing how a system of coherently coupled excited dipoles descends the Dicke ladder. At the top or bottom the decay rate is proportional to the number of emitters, $N$, but in the middle of the SF emission process it scales as $N^2$, thereby achieving advantageous ultra-short emission lifetimes via SF.

SF in QDs was first demonstrated experimentally in arrays of many lead halide perovskite QDs by Rainò et al. in 2018.[59] They showed that colloidally-synthesized lead halide perovskite QDs can be formed into micrometer sized superlattices arrays via controlled evaporation, and due to coherence effects, these QD superlattices exhibited SF at ultra-low temperatures (i.e. 6K) when excited with a 3.06 eV laser (i.e. ~405 nm) operating at 40 MHz with a pulse duration of 50 ps. Rainò et al.[59] proposed that SF-QDs could be used for entangled multi-photon quantum light sources.[59] Zhou et al., (2020)[60] demonstrated that SF from QD superlattices can be achieved at somewhat higher temperatures (e.g. 77K) using a 400 nm excitation laser (10 kHz, 40 fs pulse width) via appropriate control of the QD superlattice, which they proposed as a “quantum container”.[60] Based upon our earlier criteria, SF-QDs can be categorized as a second generation quantum technology optical nanomaterial. However, SF-QDs require cryogenic cooling, limiting their application. Notably, room-temperature SF was first demonstrated in hybrid perovskite ($CsPbBr_3$) systems in a study published in April 2022.[54] For a more comprehensive timeline of SF optical materials, readers are referred to a recent excellent review by Russ and Eisler (2024).[55]

A key technical point for room-temperature SF in lanthanide materials such as UCNPs, is that in lanthanide ions, the 4f orbitals are shielded from external environment by outer-lying occupied 5s and 5p orbitals. This means these 4f orbitals are well protected from the environment, enabling long coherence times and room-temperature SF.[52]

## 3. Experimental demonstrations of SF-UCNPs

SF-UCNPs utilize electromagnetically induced quantum coherence coupling of a large ensemble of photoactive lanthanide ions inside a UCNP into an effective single microscopic dipole, resulting in optical emission on the timescale of ns. To achieve this, a high intensity ns or fs laser pulse at an appropriate wavelength is required to induce the coherence coupling of closely packed emissive ions, by first exciting all ions into an excited state (uncorrelated dipoles of randomly orientated phases), leading to spontaneous emission and subsequent coherent coupling, generating a macroscopic giant dipole, where all dipoles are aligned in phase through as spontaneous synchronization process (see Figure 2). In the case of SF-UCNPs, the emissive ions are closely packed $Nd^{3+}$ ions, which exhibit the required properties, including robustness against dephasing due to thermal and environmental perturbation, and proclivity to $Nd^{3+}$ to $Nd^{3+}$ energy transfer via cross-relaxation processes.[52,53]

Room-temperature (i.e. 17°C/290K) SF in UCNPs was first discovered in serendipitous manner[61] and subsequently reported in a landmark paper published in *Nature Photonics* paper by Huang et al. in July 2022.[52] This research was a collaboration between the USA-based groups of Prof. Gang Han and Prof. Shuang Fang Lim of University of Massachusetts and North Carolina State University respectively. Huang et al. utilized bet-phase[62] core@shell and core@shell@shell UCNPs produced via the hot injection method and featuring a high proportion of $Nd^{3+}$ ions in either the core or shell layers (see Figure 4).[52] These UCNPs were around 50 nm in diameter and so were the smallest class of SF materials reported at that time.[61] For measurement, the UCNPs were drop cast in a dry form on a quartz glass slide (drop cast via an ethanol solution which was allowed to evaporate). SF excitation was provided by a tuneable optical parametric oscillator (OPO) and diode-pumped solid–state (DPSS) Q-switched pump laser (NT253-1K-SH-H, Ekspla), operating at 800 nm, with a pulse repetition rate of 1 kHz and a pulse width of 4.5 ns. Detection was via a single photon counting detector. The 800 nm wavelength is selected to excited $Nd^{3+}$ ions.[63–65]

Huang et al, demonstrated SF in $Nd^{3+}$ doped UCNPs where the $Nd^{3+}$ ions were localized exclusively to either the core or shell UCNP layers (see Figure 4). They noted that $Nd^{3+}$ ions have a favorable coherence state and pack closely, thereby enabling SF. They also proposed an upconversion energy transfer mechanism based upon proximal $Nd^{3+}$ ions, involving ground state absorption, cross relaxation, and excited state absorption processes. They demonstrated that a critical threshold of ~20% $Nd^{3+}$ doping was necessary to achieve SF (due to the proximity of $Nd^{3+}$ ions) and that brighter SF was observed in SF-UCNPs with a high degree of $Nd^{3+}$ doping (i.e. 90%), estimated to correspond to an average of 11 coupled $Nd^{3+}$ dipoles. With conventional continuous wave (CW) laser excitation, these UCNPs exhibited an emission lifetime of 466 μs, whereas the SF-UCNPs studied exhibited an emission lifetime of 46 ns; a ~10,000 fold improvement in emission lifetime. SF-UCNPs was observed at excitation power densities ranging from ~2 to 20 kW/cm$^2$. Huang et al., also demonstrated SF emission from both single UCNPs and clusters of UCNPs, observing variations of SF photophysics in SF-UCNP clusters, and noting that SF-UCNPs could be utilized in a wide range of applications.[52]

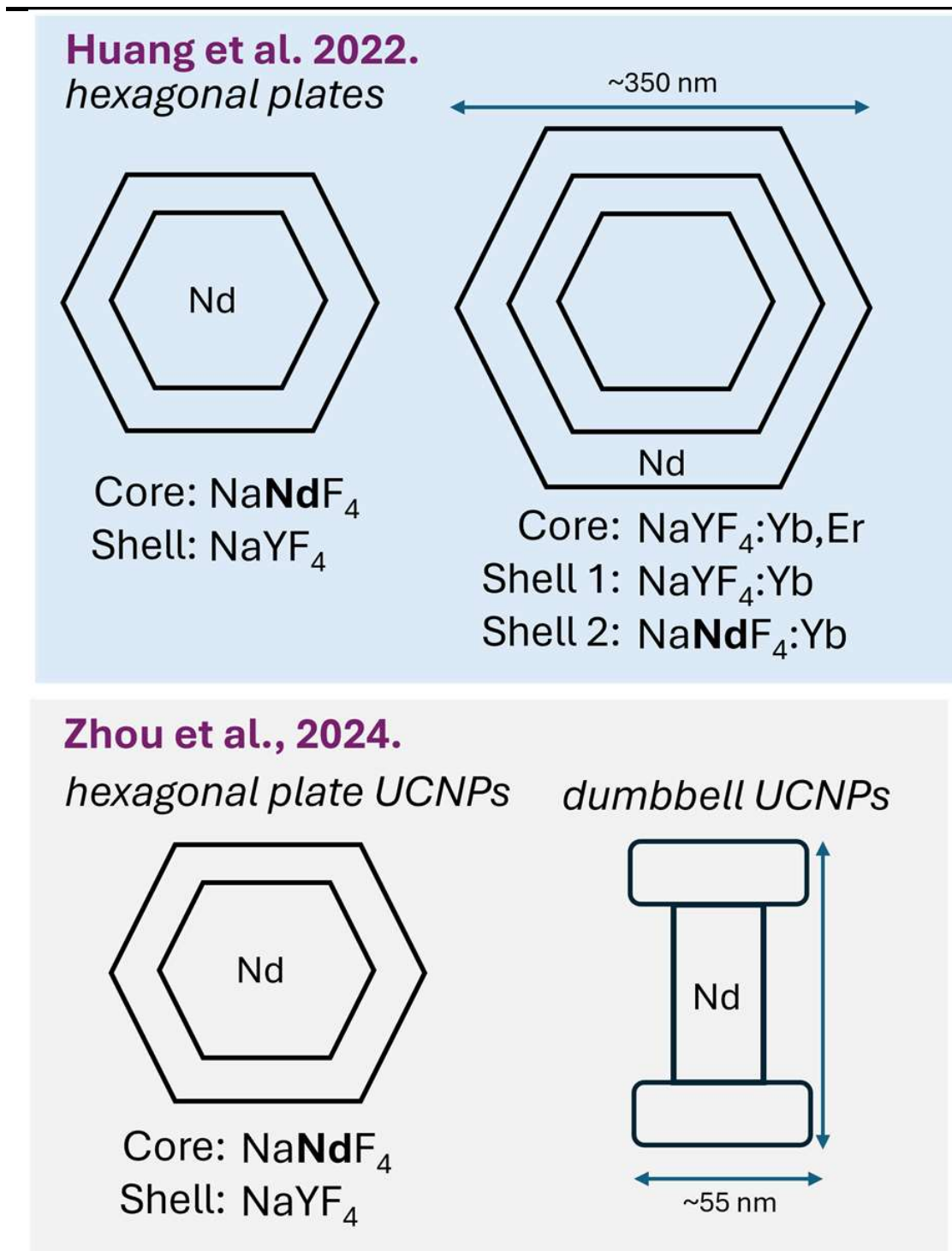

**Figure 4.** Morphologies of SF-UCNPs which have been experimentally demonstrated to-date. For clarity, only the $Nd^{3+}$ containing layers are indicated.[52,53]

A second landmark SF-UCNP was published in November 2024, by Zhou et al. in *Nature Communications*.[53] This research was led by Prof. Xueyuan Chen (China). This study built upon and substantially improved SF-UCNP performance. They used beta-phase core@shell UCNPs, produce via the hot injection method. $Nd^{3+}$ was confined to the core of the UCNPs and $Nd^{3+}$ level were varied. At low $Nd^{3+}$ doping levels, these UCNPs had hexagonal plate morphology, but at high $Nd^{3+}$ doping levels, the UCNPs were of a dumbbell morphology (i.e. exposed core with two caps of $NaYF_4$ (see Figure 4). For measurement, the UCNPs were in a dry form, having been drop-cast onto a glass coverslip via a dilute ethanol dispersion. The SF state was induced by a regeneratively amplified femtosecond Ti-sapphire laser system seeded by a femtosecond Ti-sapphire oscillator (Spitfire Pro-FIKXP, Spectra-Physics). This laser operated at a wavelength of 800 nm, a pulse repetition rate of 1 kHz, a pulse width of 120 fs, and an average energy of 4 mJ per pulse. They noted that being able to fine-tune the laser wavelength is advantageous for optimizing $Nd^{3+}$ excitation. Indeed, this has been reported in prior studies of $Nd^{3+}$ excitation in various materials.[63–65] SF-UCNP emission detection was via a photomultiplier system.

Zhou et al., found that SF was generated in all UCNPs with 25 mol% $Nd^{3+}$ or greater doping concentration, which aligns with the prior findings of a critical $Nd^{3+}$ doping level for SF by Huang et al.[52,53] Building on this, Zhou et al., found that a $Nd^{3+}$ doping level of 50 mol% in their core@shell UCNPs provided optimal SF at an excitation power density of ~2 kW/cm$^2$, postulating that this optimized coherently coupled dipole formation and dephasing. They reported that they achieved a coherent coupling of 912 $Nd^{3+}$ ions. Under conventional CW excitation, these UCNPs had an emission lifetime of ~2 μs. In their SF state, the SF-UCNPs exhibited an emission lifetime of 2.5 ns, which is an order of magnitude faster than the SF-UCNP lifetime previously demonstrated by Huang et al. (i.e. 46 ns).[52] The SF-UCNP state corresponded to a roughly 70x increase in upconversion emission intensity over conventional CW excitation (see Figure 5). Zhou et al., remarked that the number of coupled $Nd^{3+}$ ions (i.e. 912) was limited mainly by their instrumentation. Therefore they expect that it is possible to achieve a greater achieve a greater number of coherently coupled $Nd^{3+}$ ions with future improved instrumentation. Further suggestions were made by Zhou et al., such as SF-UCNPs being used for quantum optics, solid-state single-photon emission, and high-speed high rate super resolution microscopy imaging.[53]

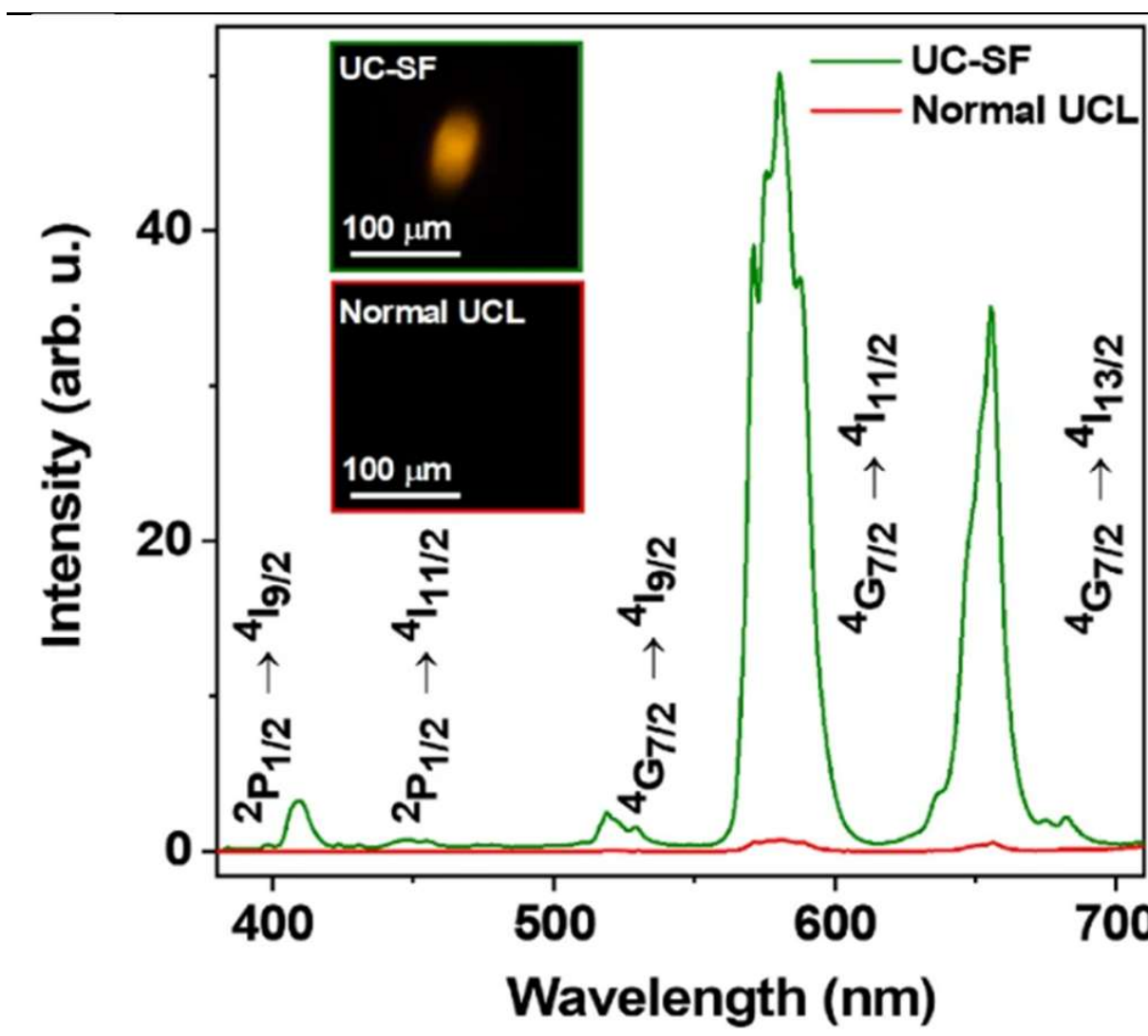

**Figure 5.** SF-UCNP demonstrated by Zhou et al., (2024)[53] demonstrating superior photon flux achieved via SF processes in comparison to conventional upconversion luminescence (UCL). 

**4. Applications of SF-UCNPs: future prospects and challenges**

The room-temperature SF-UCNPs demonstrated by Huang et al. (2022)[52] and Zhou et al. (2024)[53] are initial experimental forays into a remarkable new field which will likely attract intense interest from the upconversion nanoparticles, photophysics, and quantum physics research communities. However, applications of SF-UCNPs could cross disciplinary boundaries and move towards 2$^{nd}$ generation quantum technology applications. Some challenges of developing SF-UCNPs as a 2$^{nd}$ generation quantum technology are herein set out:

1. Fundamentally, SF-UCNP is cutting-edge and requires coordination and collaboration between: (A) skilled chemists to synthesize the sophisticated suitable SF-UCNPs; (B) experimental optical physicists and engineers (to enable optical control); (C) theoretical physicists to best understand results and propose experiments. Leadership and major international funding initiatives are required to bring together teams of researchers that can bridge these gaps and push the boundaries beyond current state-of-the-art.

2. The excitation and detection instrumentation currently required for SF-UCNP is very expensive and presents a high barrier of entry to the field. Can more affordable apparatus be found? Certainly that would expedite applications. In the meantime, SF-UCNP research provides a high barrier to entry for funding.

3. Do SF-UCNPs have to be produced via the demanding and highly skill hot-injection method, which enables generation of highly efficient core/shell UCNPs via superior control of experimental parameters such as precursor injection, heating rate, and inert atmospheres.[23] Yet, other UCNP synthesis routes are available with their own advantages and disadvantages. Could more accessible autoclave and microwave synthesis methods be utilized for production of SF-UCNPs?[22]

4. Can SF be generated in UCNP materials other than $NaNdF_4$? For example, SF effects have been demonstrated in $Er^{3+}$ doped materials at various temperatures.[66,67] Therefore, is SF emission possible in $Er^{3+}$ doped UCNPs excited at ~1500 nm?[68,69] Huang et al (2022)[52] note the dependence of SF on closely packed beta-phase unit cells (see the supplementary information of that paper for full technical discussion).[52] Given that (a) beta-phase UCNP unit cells have been achieved for other UCNP materials,[23,71] and (b) that lanthanide ions are well known to exhibit similar photophysical properties, it seems plausible that SF could be observed in UCNP materials other than $NaNdF_4$. However, SF in $NaNdF_4$ materials is contingent on interactions between neighboring $Nd^{3+}$ ions,[52] which may or may not have analogous processes in other materials. Therefore, whether or not SF can be generated in other UCNP materials requires thorough experimental and theoretical explanation of the lanthanide photophysics involved in SF.

5. What methods can be used to measure the SF quantum yield (and therefore brightness) of SF-UCNPs? This is challenging even in conventionally excited UCNPs.[44] Indeed, this Zhou et al. noted this challenge in their peer-review process.[53]

6. What effects temperature have on SF-UCNP emission? Is there a critical temperature limit to SF-UCNPs? Could SF-UCNPs be utilized for nanoscale temperature sensing with single nanoparticles?

7. Could SF-UCNPs be used for pressure sensing?[31,32] This seems plausible given the SF mechanism is dependent on close packing of $Nd^{3+}$ ions.[52] Applied pressure will vary this crystal lattice structure, potentially modulating the SF effect.

8. Plasmonic nano-cavities have been shown to induce SF in upconversion materials.[71] Could plasmonic nanomaterials likewise endow enhancements to SF-UCNPs?[33,73,74]

9. To-date SF-UCNPs have been demonstrated on a dry surface, but many applications of UCNPs are in colloidal suspension. Could SF upconversion be achieved in colloidal suspension, e.g. if the SF-UCNP material is sufficiently protected from environmental quenching?[74]

10. Are the excitation power densities required for SF-UCNPs compatible with biological applications, or are they too high/unsafe?

## Conclusions

SF in UCNPs is a room temperature quantum optical nanomaterial phenomenon which was first demonstrated in UCNPs in 2022. Whilst many fundamental aspects and practical applications of SF-UCNPs remain to be explored, the prospects of increasing UCNP emission intensity by orders of magnitude beyond conventional excitation is a tantalizing prospect that could enable SF-UCNPs to bypass the limitations of conventional UCNPs and directly compete in terms of brightness against better known high quantum efficiency materials, such as QDs. SF-UCNPs will be of interest for practical applications where higher photon flux is beneficial, such as sensing and imaging. More specifically applications such as super-resolution imaging, temperature sensing, pressure sensing, etc. However, to-date, there have only been two experimental demonstrations of SF-UCNP emission. This may reflect the expense of required instrumentation, and interdisciplinary nature of SF experiments, which is a considerable barrier to entry to the wider research community. At this early stage of their development, it is expected that SF-UCNPs can be categorized as a nascent second generation quantum optical nanomaterial.

## Competing interests

No competing interests to declare.

## Funding

L.M. had prior relevant support from the BBSRC [Discovery Fellowship] (BB/T009268/1); The Royal Society of Chemistry [Research Enablement Grant] (E21-5833576777); The Royal Society [Research Grant] (RGS\R1\221139); and The Carnegie Trust [Research Incentive Grant] (RIG013288).

## Acknowledgement

Thanks to Dr Luke McCellan for his assistance with proof-reading V1 of manuscript.